# The ALICE Silicon Strip Detector performance during the first LHC data taking

*G. Contin[1] for the ALICE Collaboration*

*(1) Università degli studi di Trieste and I.N.F.N. Sezione di Trieste*

**Abstract**

The Silicon Strip Detector (SSD) is a fundamental part of the Inner Tracking System (ITS) for the ALICE experiment. Since the early phase of proton-proton collisions at LHC, the SSD is fully operational and participating in the charged particle detection and identification carried out by ALICE. The performance of the SSD during the 900 GeV and 7 TeV collision data taking is presented here.

The stability of the system is monitored through the time evolution of its calibration parameters and their correlation with the environmental conditions. The intrinsic noise of the 2.6 million channels composing the SSD is used to assess the detector efficiency.

Finally the performances in terms of hit reconstruction and energy-loss measurement are discussed with reference to the global tracking and the ITS-standalone particle identification carried out in the first collision events.

## Introduction

ALICE (*A Large Ion Collider Experiment*) is a multi-purpose experiment designed to observe heavy ion collisions at ultra-relativistic energies at LHC. The goal of the experiment is to detect the onset and study the effects of the Quark Gluon Plasma, a high energy density and colour-deconfined state of strongly interacting matter [1, 2]. ALICE is taking data since the beginning of the LHC proton run, started in December 2009, and collected about 800 million *minimum bias* events up to October 2010. The collected data have been used for alignment and calibration and to perform benchmark measurements at the energies of $\sqrt{s}$ = 0.9 TeV and 7 TeV.

The ALICE detector consists essentially of a central part, which measures hadrons, electrons and photons, and a forward muon spectrometer. In the central part, the Inner Tracking System *(ITS)* is placed close to the interaction point and covers the pseudo-rapidity range of $|\eta| < 0.9$; a large cylindrical Time Projection Chamber *(TPC)* surrounds the ITS matching its acceptance range.

The main tasks of the ITS are to localize the primary vertices with a resolution better than 100 μm, to reconstruct the secondary vertices from hyperons and B and D meson decays, to track and identify charged particles with $p_t$ below 200 MeV/c or particles traversing the dead zones of the TPC and to improve the momentum and angle resolution for particles tracked by the TPC.

The ITS is composed of six silicon detectors layers: namely, proceeding outward from the beam pipe, the Silicon Pixel Detector *(SPD)*, the Silicon Drift Detector *(SDD)* and the Silicon Strip Detector *(SSD)*. The status and the performance of the last one, composing the two outermost layers of the ITS, is the subject of this article. Thanks to the halfway position between the innermost layer of the ITS and the internal TPC layer, the covered acceptance and the good point resolution, the SSD plays the crucial role of matching the tracks from the TPC to the ITS.

The SSD is composed of 1698 modules, each one consisting of a 1536-strip double-sided silicon sensor connected through aluminum-kapton microcables to the front-end electronics, for a total number of 2.6 million read-out channels (see also Table 1) [3]. The strip pitch (95 µm) and the relative p-n side stereo angle inclination (35 mrad) allow the SSD to provide two dimensional measurement of the track position, reaching an intrinsic point resolution of 20 µm and 800 µm in the *rφ* and *z* directions, respectively, and a good ghost point rejection. The analogue readout has a large dynamic range that allows to use the *dE/dx* measurement for particle identification in the non-relativistic *(1/β)* region.

**Table 2**: *SSD parameters*

| Radius | ~ 38 and 43 cm |
|---|---|
| Total sensitive area | (2.2 + 2.8) m² |
| Sensor area | 75 x 42 mm² |
| Number of ladders | 144 |
| Total modules | 1698 |
| Total channels | ~ 2.6 millions |

## The status of the SSD during the 2009/2010 *pp* run

The status of the SSD and its stability are well described through the calibration parameters and their time evolution: after every LHC fill injection, when the beams are not yet colliding, the SSD is calibrated through a *Pedestal Run* that collects and processes the unsuppressed data. The *Detector Algorithm* computes the pedestal, the total noise and the *common mode* corrected noise; flags a channel as *bad* if it shows large intrinsic noise *(noisy)*, extremely large pedestal *(overflow)* or constant zero output signal *(dead channel)*. The computed parameters are uploaded in the read-out electronics for the online channel masking and signal treatment and in the Offline Condition Database *(OCDB)* for the further reconstruction.

During the 2009/2010 runs, around the 92% of the SSD modules were active: 141 modules out of 1698 were not operable due to electrical and configuration problems or were masked because of noisy regions. The analysis of the OCDB objects collected in 2010 shows about 98.5% of good channels in active modules, for an overall detector efficiency of about 90.3%. As shown in Fig. 1, the bad channel fraction was stable during the Physics Runs with maximum deviations within 0.02 percentage points, except for six consecutive runs (representing ~4% of the data) where a single ladder was switched off due to a temporary problem.

The average noise is generally stable (Fig. 2*)*, although a clear dependence on the relative humidity (*RH*) of the central barrel environment affects a particular subset of ladders: when *RH* < 16%, a strong decrease of the noise is observed on the P-sides of this subset, allowing the recovery of tens of modules which would be otherwise masked in higher humidity conditions.

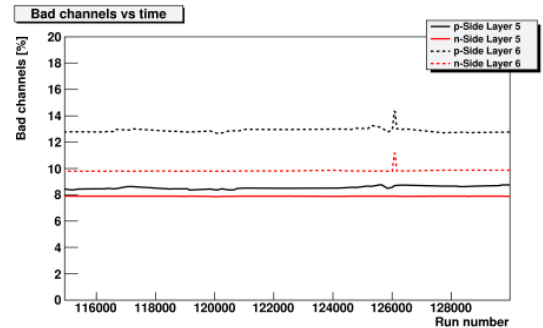

**Fig. 1**: *Time evolution of the bad channel percentage, divided by SSD module side and layer, during the 2010 LHC proton run.*

These calibration parameters are fundamental to prepare properly the detector for the physics data taking, to monitor the status of the detector on a long-term period and to describe it in the MonteCarlo simulation code.

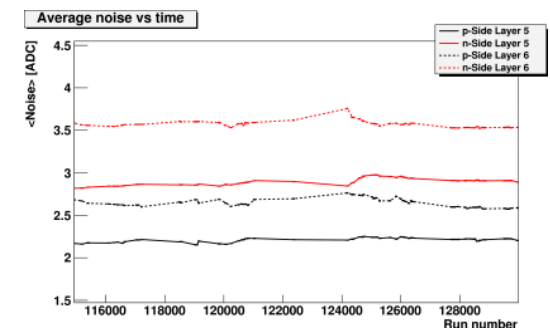

**Fig. 3**: *Time evolution of the average noise divided by module side and layer, during the 2010 LHC proton run.*

The real-time monitoring of the SSD status is carried out through the online data quality monitoring agent (*AMORE*): it provides histograms of the event data size and the occupancy, triggering a warning in case they don't match the expected ones [4]. The SSD is configured and calibrated in order to send about 32 kB per *pp* event and an average module occupancy smaller than 0.5%. These quantities are also monitored for each sector in order to localize noisy zones and to check separately the on-line data treatment: as soon as the event size increases and a module shows an occupancy higher than 3% under normal beam and run condition, it can be flagged as noisy and masked in the acquisition; on the other hand, an average occupancy unexpectedly equal to zero can indicate a misconfiguration of the module.

At each variation of the multiplicity conditions in ALICE, the signal processing performed by the acquisition hardware is assessed, especially for what concerns the compensation of the shift generated by the *Common Mode (CM)* noise, namely the noise component influencing coherently all the 128 channels connected to the same front-end chip. For each event, the baseline shift is calculated as the average of the signals collected in the channels not fired by particles, and subtracted. The correction is monitored through the CM distribution per chip and the global distributions of the means and the standard deviations (see Fig. 3). The chips showing too large CM mean or RMS are masked: the CM correction is not reliable for them and produces an artificial enhancement or suppression of the particle signal amplitudes and a modification of their original cluster shape. In addition, CM distribution for each chip is monitored to spot correction failures due to baseline drifting or high noise zones.

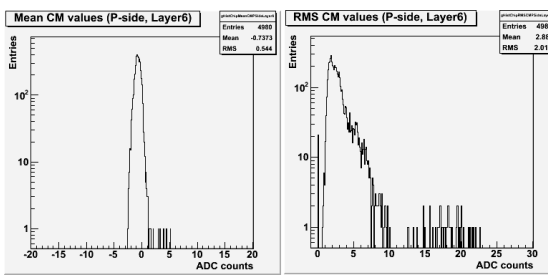

*Fig. 3*: *Distribution of the Common Mode means (left) and standard deviations (right) for the SSD chips in Layer 6 p-sides.*

After the reconstruction process the data quality is monitored at different stages with dedicated tools. Firstly the *Quality Assurance* analysis is automatically performed after the reconstruction of each run. The distribution of the cluster charge per module (Fig. 4) is used to calibrate the gain at the module level: after this equalisation, the most probable values *(MPVs)* of the charge distributions of the charge collected by each module are stable within a few percent. The results are also used for the update and optimisation of the OCDB *Gain* object, where the signal amplitude gain factor is stored for further reconstruction passes.

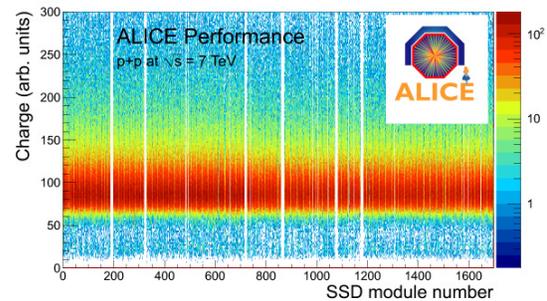

*Fig. 4*: *dE/dx in SSD modules as a function of module number for pp at 7 TeV. The color scale represents the number of entries in a given bin.*

A comparison of the space point distributions in simulated and real data is performed offline. The *MonteCarlo/Data* ratio distribution per module allows to estimate the reconstruction efficiency and to improve the description of the detector response in the simulation code. This analysis is able to spot particular defects, like for example modules highly inefficient in collecting and amplifying charge and modules with noise strictly correlated to the humidity. At present, the consequent tuning of the masking thresholds allows a proper description of the 99% of the modules in the code.

## The SSD performance during the 2009/2010 *pp* data

After the accurate commissioning activities, the SSD has been included in physics acquisition since the very first collisions in December 2009 (0.9 TeV) and in March 2010 at the nominal energy of 7 TeV. Thanks to the stable conditions and the overall efficiency maintained along the whole running period, it took part in almost all the physics runs and collected about $1 \cdot 10^9$ events in about 1000 hours of data taking.

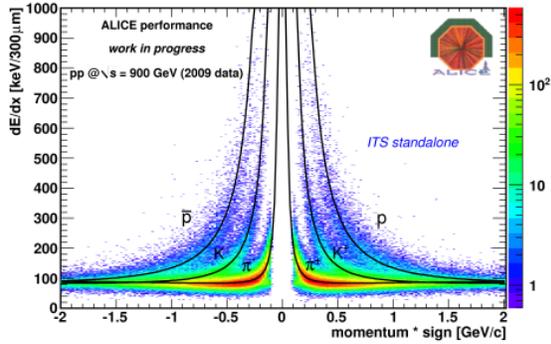

**Fig. 5**: *SDD+SSD dE/dx versus track momentum (ITS standalone) shown for particles and antiparticles separately.*

The performance of the SSD during the data taking in terms of point resolution and charge measurement has been qualified by the analysis of the collision data.

The study of SSD alignment through the double points in acceptance overlaps shows a sigma of the residuals compatible with the expected intrinsic single point resolution of 20 μm in the *rφ* direction.

The SSD provides also the measurement of the charged particle energy loss in the detector. These values together with the SDD measurements, once corrected for the path length, are used to identify π, K, p tracked by the ITS standalone (low $p_t$) or together with the TPC. The *dE/dx* is calculated with the truncated mean method, namely calculating the mean of the two lowest normalized charge signals collected by the SDD and SSD layers. The measured energy loss as function of the total momentum is shown in Fig. 5 for the mentioned (anti-)particles.

## Conclusions

The ALICE Silicon Strip Detector is participating successfully in the ALICE data taking since the beginning of the LHC *pp* run.

Its overall efficiency larger than 90% and its good stability allow the SSD to provide the expected performance in terms of space point resolution and *dE/dx* measurements for particle identification.

The detector is now fully commissioned and its response is completely understood: therefore the system is in a good shape for the upcoming heavy ion collisions.

## References

[1] ALICE Collaboration *et al.* (2004): *ALICE: Physics Performance Report, Volume I, J. Phys. G.,30, pp.1517-1763*

[2] ALICE Collaboration *et al.* (2006): *ALICE: Physics Performance Report, Volume II, J. Phys. G.,32, pp.1295-2040*

[3] ALICE Collaboration *et al.* (2008)*: The ALICE experiment at the CERN LHC, Jinst 3, S08002*

[4] P. Christakoglou, I. Kielen (2009)*: Online Data Quality Monitoring for the ALICE Silicon Strip Detector, ALICE Internal Note*